\documentclass[12pt, letterpaper]{article}

\usepackage{framed,multirow}

\usepackage{amsmath,amsfonts,amssymb}
\usepackage{latexsym}
\usepackage{amssymb}
\usepackage{booktabs}
\usepackage[export]{adjustbox}
\usepackage{url}
\usepackage{xcolor}
\usepackage{todonotes}
\definecolor{newcolor}{rgb}{.8,.349,.1}

\usepackage{hyperref}
\usepackage[capitalize]{cleveref}
\crefname{section}{Sec.}{Secs.}
\Crefname{section}{Section}{Sections}
\Crefname{table}{Table}{Tables}
\crefname{table}{Tab.}{Tabs.}

\usepackage{pifont}

\usepackage[switch,pagewise]{lineno}

\title{ScanMove: Motion Prediction and Transfer for Unregistered Body Meshes}

\author{Thomas Besnier$^1$, Sylvain Arguill\`ere$^2$   Mohamed Daoudi$^3$ $^,$ $^4$}
\date{%
    $^1$Univ. Lille, CNRS, Centrale Lille, UMR 9189 CRIStAL, Lille, F-59000, France\\%
    $^2$Univ. Lille, CNRS, UMR 8524 - Laboratoire Paul Painlev\'e, F-59000 Lille, France \\
     $^3$Univ. Lille, CNRS, Centrale Lille, Institut Mines-Télécom, UMR 9189 CRIStAL, Lille, F-59000, France\\
     $^4$IMT Nord Europe, Institut Mines-Télécom, Univ. Lille, Centre for Digital Systems, Lille, F-59000, France\\[2ex]
}

\begin{document}

\maketitle










\begin{abstract}
Unregistered surface meshes, especially raw 3D scans, present significant challenges for automatic computation of plausible deformations due to the lack of established point-wise correspondences and the presence of noise in the data. In this paper, we propose a new, rig-free, data-driven framework for motion prediction and transfer on such body meshes. Our method couples a robust motion embedding network with a learned per-vertex feature field to generate a spatio-temporal deformation field, which drives the mesh deformation. Extensive evaluations, including quantitative benchmarks and qualitative visuals on tasks such as walking and running, demonstrate the effectiveness and versatility of our approach on challenging unregistered meshes.
\end{abstract}











\begin{figure*}[ht!]
    \centering
    \includegraphics[width=1.0\linewidth]{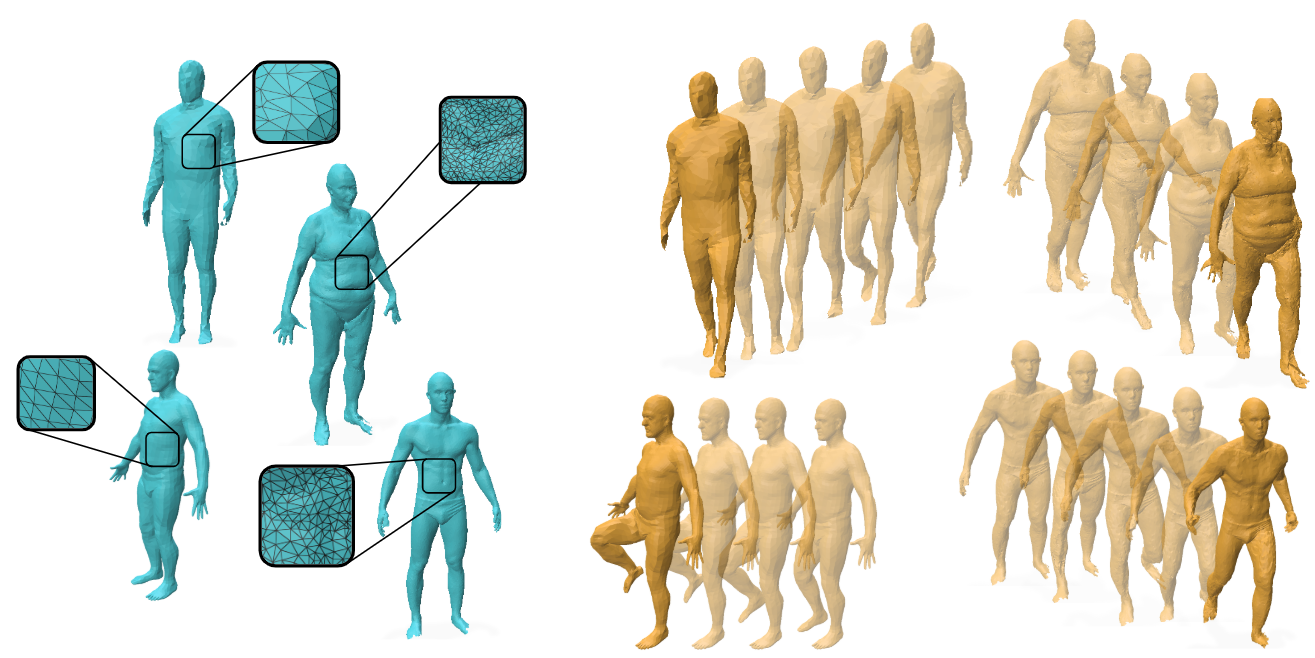}
    \caption{\textbf{ScanMove} is a learning-based framework designed to automatically predict deformations of unregistered surface meshes. From an arbitrary unregistered triangulated surface of a body, including scanned data (\textcolor{cyan}{on the left}), ScanMove can predict complex and long-range motions as time-dependent deformations of the input mesh (\textcolor{orange}{on the right}).}
    \label{fig:intro_figure}
\end{figure*}

\section{Introduction}
\label{sec:intro}

In 3D computer vision, deformation modeling has seen tremendous improvements regarding human body movement generation and behavior prediction. Several models and frameworks are now available to tackle complex human body motions such as walking, running, or even dancing. However, most of these methods are built upon a parametric model such as the widely appreciated SMPL body model \cite{SMPL:2015, STAR:2020}. The latter offers a mapping from a 3D rigid skeleton, the \textit{rig}, characterized by a specific set of joints, to a  triangular, fixed topology surface mesh: this mapping from the rig to the surface mesh is called a \textit{skinning} map. The limitation comes when unregistered data such as scans or challenging synthetic surface meshes are presented (see some examples on the left in~\cref{fig:intro_figure}). Recent advances in 3D scanning technologies have enabled the capture of highly detailed and complex body meshes in uncontrolled environments. However, these "in-the-wild" scans often present noise, irregular sampling, and lack of registration, posing significant challenges for traditional rig-based animation techniques. Our work aims to address this issue by introducing a robust, rig-free framework capable of predicting realistic and temporally coherent deformations directly from raw mesh data. This capability not only enhances the flexibility of motion synthesis but also opens new avenues for applications in augmented reality, virtual character animation, and robotics. A straightforward approach can be to register this new data by optimizing the pose and identity parameters. However, without careful initialization and proper data acquisition, this approach can be both costly and may struggle to capture fine-grained details. Moreover, parametric models are limited to a category of shapes (e.g. human bodies for SMPL) and require additional work to adapt to new categories. Our goal is to propose a generic method for any category of non-rigid surface mesh data. Additionally, the proposed framework is not limited to human motions as shown later in~\cref{fig:motion_transfer_ex}.

Most existing methods necessitate additional explicit motion priors such as joint position, especially during training. In this work, we present a recursive model that takes as input any source mesh and unregistered motion sequence and generates a complex motion as a time-dependent and vertex-wise deformation field, overcoming several previous limitations. 
Our method does not require any skeletal information (for training and inference) and to be versatile enough for complex motions such as walking and running and robust enough to be used in raw scans, as shown in~\cref{fig:intro_figure}. In comparison, SMPL based methods are highly efficient but require rigging, and skinning and cannot be used to deform unregistered meshes. Deformation models such as \cite{groueix2018_3DCODED, NJF_2022} are robust and versatile but cannot directly capture complex time-dependencies. 
Recent approaches are breaking free of the need for a consistent skinning mapping~\cite{TRJ, riganything2025} but heavily rely on a good rigging. In this work, we suppress the need for any skeletal rig and encode the motion with a simple yet effective robust embedding network.

To summarize, the contributions presented in this paper consist in \emph{(i)} a new end-to-end deep learning framework for deformation prediction to predict complex motions; \emph{(ii)} we detail and explore the learned space-time feature space to predict vertex-level deformations; \emph{(iii)} comprehensive comparisons and visualization of applications such as motion prediction and transfer on challenging unregistered meshes like raw scans.

\section{Related work}
\label{sec:related_work} 

\subsection{Triangular mesh deformation}

Computing surface mesh deformations is a long standing challenge in computer graphics. Most methods can be either defined on the surface itself or acting on the ambient space (see \cite{polygon_mesh_processing}, chap. 9 for more). The former includes intrinsic energy minimizing optimization (shell \cite{Terzopoulos_1987, Celniker_1991}, elastic \cite{Hartman_H2_elastic_match, BaRe-ESA_2023_ICCV}, ARAP \cite{ARAP_2000, ARAP_2007}, and others \cite{PriMo_2006}), multi-scale methods \cite{Botsch_Sorkine_2008}, vector displacements \cite{DeRose_1998, Guskov_1999} and differential coordinates editing (gradient based \cite{Yu_2004, Zayer_2005} or laplacian based \cite{Lipman_2008, Sorkine_2004}). However, because these methods are essentially linear, they tend to fail for large deformations such as complex human motions. Also, they often involves solving complex optimization processes and post processing (local correction, smoothing, ...) which can be time consuming for high resolution meshes such as scans. In contrast space deformations are fully extrinsic and include in particular lattice-based deformations \cite{Sederberg_Parry_1986} and cage-based deformations \cite{Thiery_2018, Ju_2005, Ju_2007, survey_cage-based_deformation_2024}. However, these methods are also challenging to use for large non-linear deformations and generalized harmonics or Green functions can induce undesirable non-local changes. Recent learning-based models aims at improving the generalizability and flexibility of such methods \cite{Yifan_NeuralCage_2020} but still suffer from similar drawbacks. In contrast, we propose to leverage recent deep neural deformation model arranged in a new recursive framework designed to tackle dynamic motions and non linear deformations by learning the latter from data.

\subsection{Mesh-Invariant Learning of Deformations}
Existing methods for mesh deformations that are said to be "robust" against remeshing fall into two categories: 1) methods that animate a fixed template mesh to match an unregistered target shape.  2)  methods that predict per-point (or per-vertex/face) deformations directly on an unregistered mesh.
The framework presented in this paper belongs to the second category, which can produce "in-the-wild" deformations. To achieve this, recent geometric deep learning research works proposed acting on a vertex level \cite{groueix2018_3DCODED, sharp2021diffusionnet, attaiki2023snk, SMS_2024, NJF_2022} to avoid the connectivity constraint. Some recent works that bear similarities with our approach include methods that combine spectral and spatial representations to obtain mesh-invariant features \cite{SMS_2024} as well as frameworks relying on Riemannian shape analysis \cite{BaRe-ESA_2023_ICCV}. Building on this, Temporal Residual Jacobians (TRJ) \cite{TRJ} couples spatial deformation predictors with neural ordinary differential equations (ODEs) to ensure temporal coherence over long motion sequences. Previous works have addressed localized deformations in specific contexts, such as facial expression synthesis \cite{NFR_2023} and speech-driven lip motion synthesis \cite{nocentini2024scantalk3dtalkingheads}) but relatively few have successfully tackled full-body motion without any rigging \cite{TRJ}, especially when considering long-range dynamics like walking or running.\\ 
Our proposed method builds on these recent ideas by further decoupling the mesh deformation process from any fixed skeletal structure. By directly predicting per-point deformations on unregistered meshes—and by leveraging both spatial and temporal cues, our framework achieves robust motion synthesis even in the presence of topological inconsistencies and scanning noise.

\subsection{Human Motion Synthesis and prediction}
\begin{figure}[ht!]
    \centering
    \includegraphics[width=0.9\linewidth]{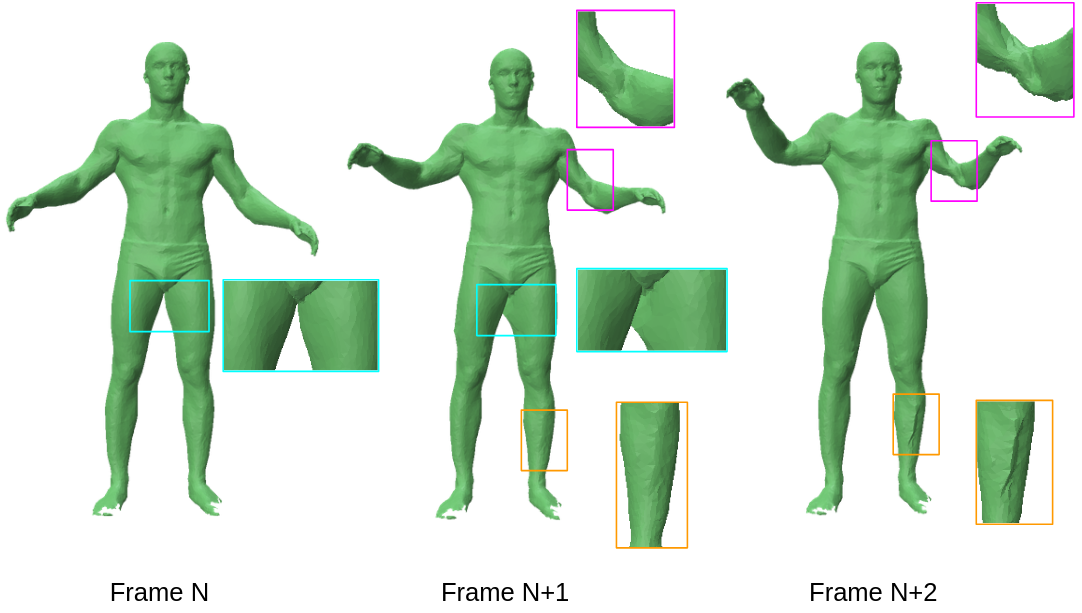}
    \caption{Jittering artifact when transferring motions using \cite{deformation_transfer_2004}. We used the python implementation of the algorithm \protect\footnotemark.}
    \label{fig:jittering_2004}
\end{figure}

Human motion synthesis can be naturally formulated as a sequential deformation prediction problem. Because the space of plausible motions is high dimensional and highly non-linear, early approaches relied on skeletal rigging and skinning methods \cite{SCAPE, SMPL:2015, STAR:2020}, in which a fixed skeleton is mapped onto a template mesh with pre-defined connectivity. Especially, the motion is entirely conditioned on this predefined skeleton which allows one to generate complex motions \cite{guo2020action2motion, petrovich21actor, xu2023actformer, diomataris2024wandr, correspondence_free_motion_retargeting_2024} from a set of rigid joint rotations. This strong control allows one to condition the generated motion on multi-modal inputs such as text description \cite{tevet2022motionclip, MotionDiffuse2024, zhang2025motion_mamba}, audio files \cite{Huang2020DanceRL} and another avatar behavior \cite{ghosh2024remos}. Although these rig-based approaches offer efficiency and precise control, they require all meshes to conform to a fixed topology and depend on a manually designed skeleton, limiting the flexibility in applications where scans are unregistered or exhibit variable connectivity. In parallel, unrigged, static deformation transfer has been tackled \cite{song20213d, wang2020neural, wang2023posetransfer, skeleton-free_2022, deformation_transfer_2004} but has not been extended to dynamic settings yet and may show jittering artifacts because of the lack of time consistency optimization as shown in~\cref{fig:jittering_2004}.\\
\footnotetext{\url{https://github.com/moChen0607/Deformation-Transfer-for-Triangle-Meshes-1}}

Recent advances in geometric deep learning have begun to relax these constraints. Instead of imposing a rigid prior on the underlying motion, several methods now directly learn deformation fields on the mesh using graph convolutional neural network (GCN) \cite{chebnet_2016, Meshconv_2020, verma2018feastnet}. These enable us to build efficient neural 3D morphable models \cite{bouritsas2019neural3dmm} with applications to motion predictions \cite{LEMEUNIER2022131} but the latter are limited to registered data.\\

\begin{figure*}
    \centering
    \includegraphics[width=0.9\linewidth]{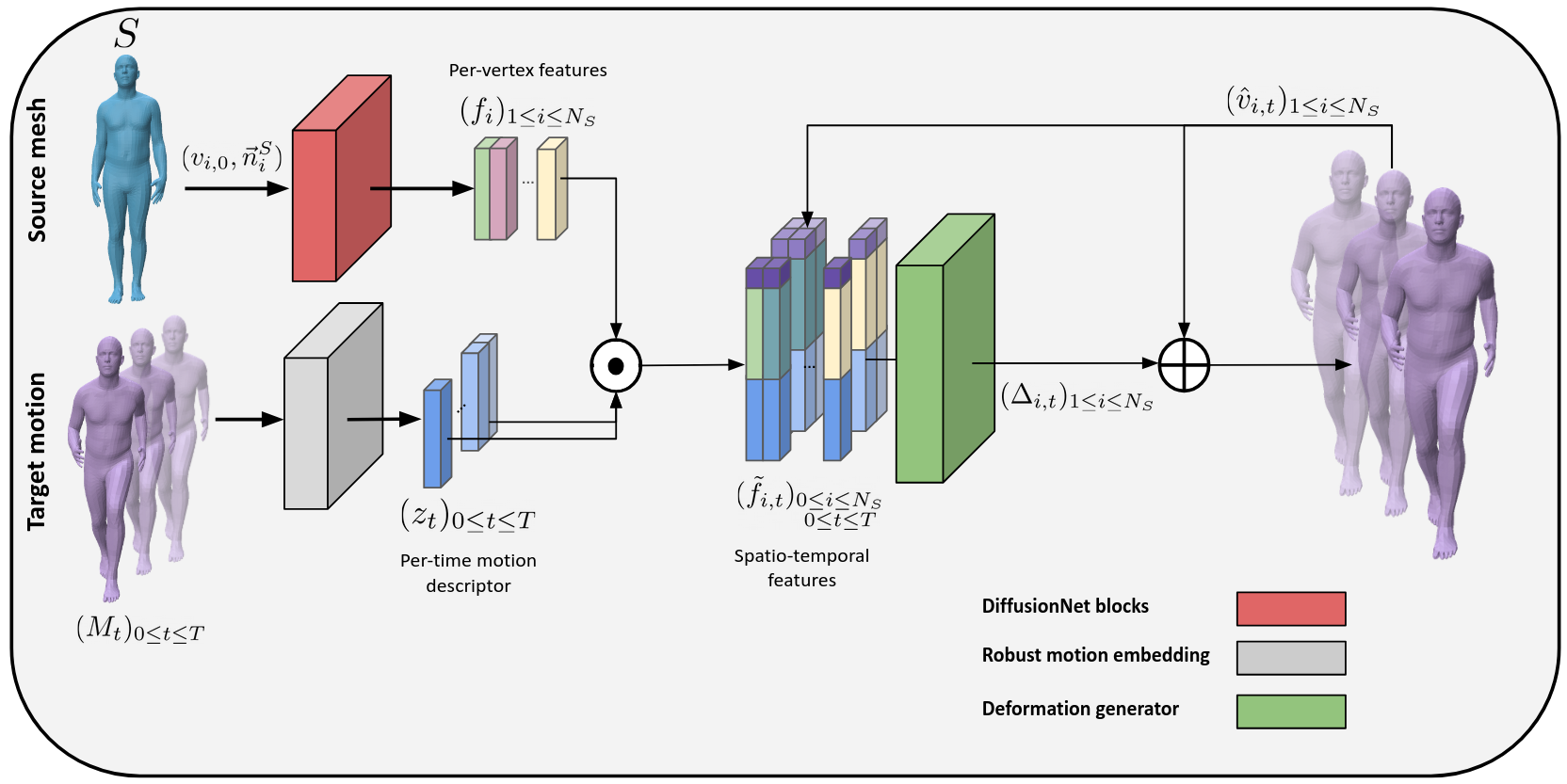}
    \caption{\textbf{Overview of the proposed framework.} From vertex positions and normals, ScanMove predict per-vertex features. In parallel, the target motion is encoded as a smooth vector time series. Concatenating these gives a 3D tensor of spatio-temporal features. From these features, a recursive multi-layer perceptron predict the successive deformations of the source mesh so it matches the target motion.}
    \label{fig:model_overview}
\end{figure*}

\section{Method}
\label{sec:method}

Given a triangulated source mesh $S=(\mathcal{V}, \mathcal{T})$ such that $\mathcal{V} = (v_{i})_{0 \leq i \leq N_S}$ is its (arbitrary) set of vertices and $\mathcal{T}$ is its face connectivity, often referred as the \textit{topology} of the mesh. Our aim is to animate $S$ given a motion sequence $(M_t)_{1\leq t \leq T}$ of arbitrary length $T>0$. Effectively, we want to predict the vertex trajectories $(v_{i,t})_{\substack{0 \leq i \leq N_S \\ 0 \leq t \leq T}}$.
In our setting, $S$ can have an arbitrary arrangement of its triangles and an arbitrary resolution $N_S$, referring here to the number of vertices. Additionally, the meshes in $(M_t)_{1\leq t \leq T}$ are also not aligned with any fixed topology and no skeletal rig is available a priori. \\

\subsection{Overview}
First, ScanMove predicts a vertex-wise feature field on $S$, with a \textbf{robust mesh feature extractor}. Simultaneously, we embed the motion sequence as a vector time series using a \textbf{robust motion embedding network}. Then, for each time step $t$, the corresponding motion vector is concatenated to each feature on the feature field, along with the vertex positions at time $t-1$ to form a new augmented feature field. This new feature field is used as input signal to a \textbf{deformation generator} predicting a deformation field that is added to the source mesh $S$. An overview of the proposed model is presented in~\cref{fig:model_overview}.

\subsection{Robust mesh feature extractor.} 
To predict a vertex-wise feature field, we feed vertex positions $(v_i)_{1 \leq i \leq N_S}$ and normals $(\Vec{n}_i)_{1 \leq i \leq N_S}$ of the source mesh $S$, to 4 DiffusionNet blocks \cite{sharp2021diffusionnet}. This deep learning network relies on heat diffusion over surfaces (through a learned diffusion time) to derive a robust convolution operation to extract discretization-agnostic features based on the local geometry of the mesh. In our framework, the feature extractor is denoted as $DN$ such that \\
\begin{equation}
    DN(S) = (f_i)_{1 \leq i \leq N_S} \quad \in \mathbb{R}^{N_S \times F}
\end{equation}
where $F$ is the feature dimension.\\
This feature extractor is robust to a change of mesh topology so it focus on learning local geometric features. It is robust to changes in resolution, face connectivity and can also be adapted on point clouds \cite{sharp2021diffusionnet} . Hence, the source mesh can have any resolution $N_S$ and can generalize to unregistered scans as shown experimentally in \cite{NFR_2023, LJN_2024}. The limitation comes when the mesh presents disconnected components, in which case the intrinsic heat diffusion process cannot reach the other components.

\subsection{Robust motion embedding network.} 
\begin{figure}[ht!]
    \centering
    \includegraphics[width=0.9\linewidth]{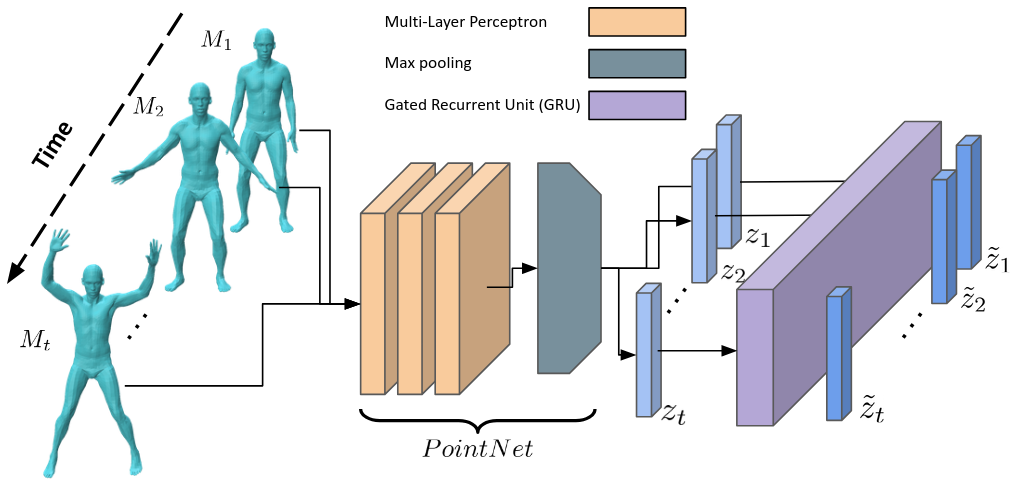}
    \caption{\textbf{Motion embedding network.} From an unregistered sequence of meshes, we encode each frame with a robust encoder coupled with 3 layers of GRU for temporal consistency.}
    \label{fig:motion_embedding_details}
\end{figure}

In order to guide the deformation of the source mesh in a time consistent way, we encode a time-dependent descriptor for the target motion with a motion embedding network detailed in~\cref{fig:motion_embedding_details}. For each frame $M_t$ of the motion sequence, we map the mesh to a latent vector $z_t$ of size $d$ using a PointNet \cite{Charles_PointNet_2017} encoder for which the code is taken from the official implementation of \cite{Achlioptas2017LearningRA}. Doing this for each frame yields a collection of latent vectors $(z_t)_{1 \leq t \leq T}$. Then, this collection of vectors is fed to three layers of bi-directional Gated Recurrent Units (GRU) to improve time coherence. Hence, we obtain a vector time series $(\Tilde{z}_t)_{1 \leq t \leq T}$ that describe the motion. \\
\begin{equation}
    \Tilde{z}_t = GRU(PointNet(M_t)) \quad 0 \leq t \leq T
\end{equation}

Note that this motion embedding network can be used to learn a "dictionary" of complex movements, similar to what audio encoders do with speeches. With more motions in the training data, more movements are accessible to the deformation decoder. More details are given in~\cref{subsec:applications}.

\subsection{Time-dependent deformation generator.}
At a given (discrete) time step $t$, we duplicate and concatenate the corresponding motion vector $\Tilde{z}_t$ along with the predicted vertex position of the previous time step $\hat{v}_{i,t-1}$ to each features $f_i$ to obtain a new feature field $\Tilde{f}_{i, t}$ such that
\begin{equation}
    \Tilde{f}_{i, t} = (f_i, z_t, \hat{v}_{i,t-1}) \quad \in \mathbb{R}^{N_S \times (f + d + 3)}
\end{equation}

\noindent Effectively, the space-time feature tensor $(f_{i, t})_{\substack{0 \leq i \leq N_S \\ 0 \leq t \leq T}}$ is a 3D tensor. This augmented feature field now has the information of how to deform each point of the source mesh at time $t$ and so, for each time.\\
Finally, at time $t>0$, the deformation generator $D$, which is a shared MLP, predicts the displacement field $(\Delta_{i,t})_{1 \leq i \leq N_S}$ from the the predicted vertex position at time $t-1$ so that
\begin{equation}
    v_{i,t} = \Delta_{i,t} + v_{i,t-1} = D(\Tilde{f}_{i, t}) + v_{i,t-1}
\end{equation}
To wrap up, ScanMove learns a spatio-temporal deformation field $(\Delta_{i,t})_{\substack{0 \leq i \leq N_S \\ 0 \leq t \leq T}}$ independently of $T$ and $N_S$. 

\subsection{Training strategy.} \label{subsec:training_strategy}
\textbf{Registered setting.} In this setting, the framework is trained end-to-end and the optimization can be supervised with ground truth motions. Hence, we use the Mean Squared Error (MSE) loss $\mathcal{L}^{MSE}$ coupled with a cosine similarity loss $\mathcal{L}^{N}$ between predicted normals and ground truth normals for the reconstruction loss so that $\mathcal{L}^{rec} = \mathcal{L}^{MSE} + \lambda_N\mathcal{L}^{N}$. Additionally, to improve the smoothness of predicted sequences, we add an As-Isometric-As-Possible (AIAP) regularization loss $\mathcal{L}^{I}$ based on \cite{aiap_2007} and used in \cite{groueix2018_3DCODED} to avoid small non-isometric local deformations. \\
\begin{equation}
    \mathcal{L} = \mathcal{L}^{MSE} + \lambda_N\mathcal{L}^{N} +  \lambda_I\mathcal{L}^{AIAP}
\end{equation}

Then, if we note $\hat{M}_t$ the predicted mesh at time $t$,
\begin{equation}
    \mathcal{L}^{MSE}(\hat{M}_t, M_t) = \frac{1}{T.N_S} \sum_{t=0}^T \sum_{i=1}^{N_S} \|v_{i,t} - \hat{v}_{i,t} \|_2^2
\end{equation}
where $v_{i,t}$ is the ground truth vertex position at time $t$ and $\hat{v}_{i,t}$ its predicted counterpart and

\begin{equation}
    \mathcal{L}^{N} = \frac{1}{T.N_S} \sum_{t=0}^T \sum_{i=1}^{N_S}\left( 1 - \Vec{n}_{i}^{M_t}.\Vec{n}_{i}^{\hat{M}_t} \right)
\end{equation}
where $\Vec{n}_{i}^{M_t}$ is the ground truth normal at vertex $v_i$ and time $t$ and $\Vec{n}_{i}^{\hat{M}_t}$ its predicted counterpart.\\
For the regularization loss $\mathcal{L}^{I}$, it can be written as \\
\begin{equation}
    \mathcal{L}^{I} = \frac{1}{T|\mathcal{E}|} \sum_{t=0}^T\sum_{(i,j)\in \mathcal{E}} \left( \frac{\|\hat{v}_{i,t} - \hat{v}_{j,t}\| - \|v_{i,0} - v_{j,0}\|}{\|v_{i,0} - v_{j,0}\|} \right)^2
\end{equation}
where $\mathcal{E}$ is the set of edge indices.\\
Also, note that we observed better results when adaptively adding the regularization after several epochs. Details will be given in the next section.

\begin{figure*}[ht!]
    \centering
    \includegraphics[width=1.0\linewidth]{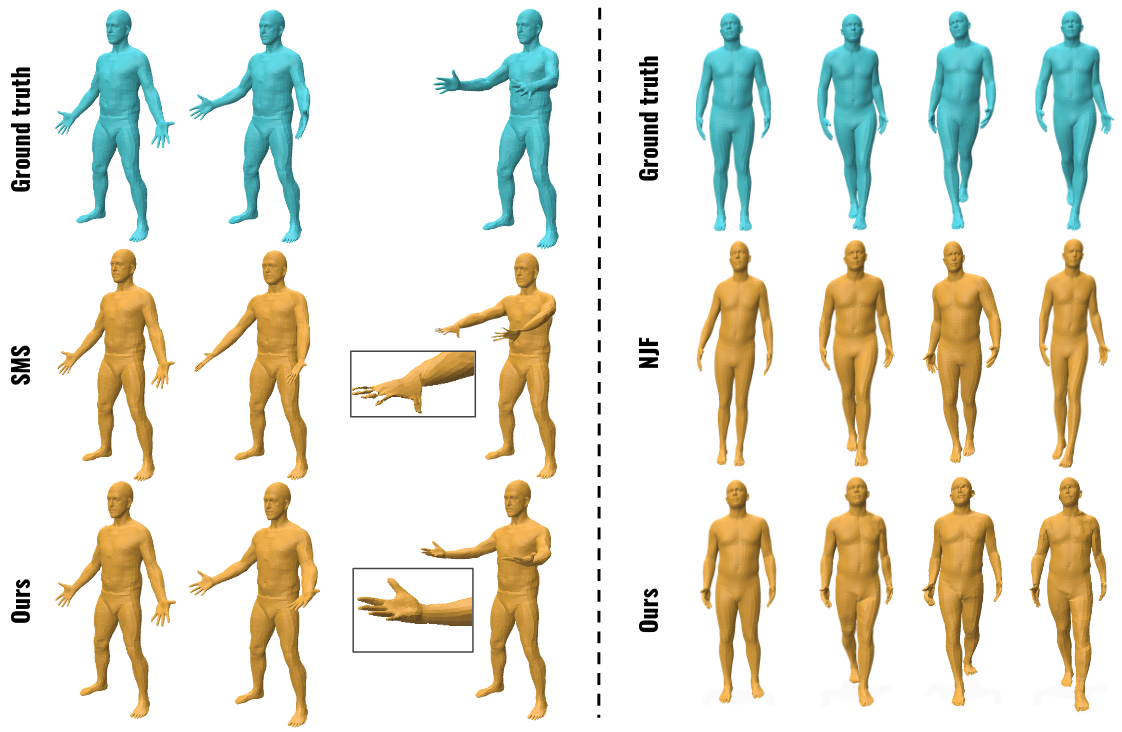}
    \caption{\textbf{Qualitative comparisons for DFAUST and WALK on a registered setting.} We display qualitative comparisons for a deformation prediction task against contestants that showed the best quantitative results. We observe wrong deformations produced by SMS on regions with irregular arrangement of the triangles such as the hands.}
    \label{fig:qualitative_comparisons_DFAUST}
\end{figure*}

\noindent \textbf{Unregistered training.} Interestingly, the framework can be trained in a complete unsupervised setting in which the vertex-wise correspondence of each mesh in the training set is unavailable. For this, we rely on the chamfer distance (CD) for the reconstruction loss so that $\mathcal{L}^{rec} = \mathcal{L}^{CD}$ with 

\begin{multline}
    d_{CD}(\hat{M}_t, M_t) = \frac{1}{N_S}\sum_{i=1}^{N_S}\min_{j}\|\hat{v}_{i,t} - v_{j,t} \|_2^2 + \\ \frac{1}{N_t}\sum_{j=1}^{N_t}\min_{i}\|v_{j,t} - \hat{v}_{i,t}\|^2_2
\end{multline}
so that 
\begin{equation}
    \mathcal{L}^{CD} =\frac{1}{T} \sum_{t=0}^T d_{CD}(\hat{M}_t, M_t) 
\end{equation}

It is worth noting that, to our knowledge, this is the first formulation of complete unsupervised prediction of complex motions for unregistered meshes.

\section{Experiments}
\label{sec:experiments}

\subsection{Experiments setup}
\label{subsec:setup}

\textbf{Datasets.} For movements with large deformations in place, we rely on the \textit{DFAUST} dataset \cite{dfaust:CVPR:2017}. The original dataset contains many redundant shapes, so we resampled it to have one elementary movement for each sequence. The final dataset comprises 10 human subjects that perform between 3 and 5 simple movements each. \\
For longer range motions, we use an augmented version of the walking motions of \textit{AMASS} \cite{AMASS_ACCAD} (ACCAD). From the original motions, we randomly modified the identity to improve diversity. The final dataset contains 7 identities with 3 walking motions, each in a different direction. We call this dataset \textit{WALK}. \\
More details on the datasets are given in~\cref{tab:datasets_specs}. Note that ScanMove operates with relatively few training samples compared to most other methods.

\begin{table}[ht!]
\centering
\begin{tabular}{l@{\hspace{0.5cm}}l@{\hspace{0.3cm}}l@{\hspace{0.3cm}}l}
\toprule
         & \textbf{DFAUST} & \textbf{WALK}  \\ 
\midrule
Type                 & In place & Long-range \\
\# vertices          & 6890 & 6890  \\
\# faces             & 13776 & 13776  \\
\# meshes (training)     & 325 & 615  \\ 
\# sequences (training)  & 35 & 15  \\
\# meshes (test)       & 61 & 246  \\
\# sequences (test)    &  7 & 6  \\
\bottomrule
\end{tabular}
\caption{\textbf{Train / test splits for each dataset.}}
\label{tab:datasets_specs}
\end{table}

\noindent \textbf{Metrics.} We report the vertex-wise mean square error (MSE) between predicted and ground-truth sequences. \\
We also report the mean deviation between the predicted and ground-truth normal maps using a mean cosine dissimilarity (Cosim). The formulation of these metrics is similar to the corresponding loss functions introduced in~\cref{subsec:training_strategy}.

\noindent \textbf{Implementation details.} All models are trained for 500 epochs using the Adam \cite{ADAM_2014} optimizer. The learning rate starts from $10^{-3}$ and is progressively reduced with a Step LR scheduler of step size 5 and a decay rate of 0.99. We set $\lambda_N=10^{-4}$ and $\lambda_I=10^{-5}$ with $\lambda_I$ added after 400 epochs. Training and inferences were performed on a computer equipped with an Intel Xeon Gold 5218R CPU with 64Go RAM and a NVIDIA Quadro RTX 6000 graphic card.

\noindent The feature extractor is composed of 4 DiffusionNet \cite{sharp2021diffusionnet} blocks predicting a per-vertex feature vector of size $f=64$. The motion embedding network is a shared multi-layer perceptron with 4 layers of size 128 with LeakyReLU activations followed by a maxpool aggregation to obtain a latent code of size $d=64$ (as a lightweight PoinNet encoder \cite{Charles_PointNet_2017}) and end with 3 layers of bidirectional GRU to predict a vector time series of size $T \times d$ where $T$ is the arbitrary sequence length. The recursive deformation generator is a shallow MLP, shared by all vertices, with 4 layers of size 128 with LeakyReLU activations. 

\subsection{Large deformations in place}

In this experiment, we evaluate the deformation predictions on DFAUST. The test set comprises two unseen identities executing similar (but not exactly the same) movements as seen in the training set. We compare the performance of ScanMove with other recent deformation models. Meshconv \cite{Meshconv_2020} and ArapReg \cite{ARAPReg} are convolutional graph neural networks. Note that these models cannot, by design, deform unregistered meshes. Neural Jacobian fields (NJF) \cite{NJF_2022} predict the Jacobian of the deformation and reconstruct the vertex positions of the deformed mesh by solving a Poisson equation. Finally, Spectral Meets Spatial (SMS) \cite{SMS_2024} is a shape interpolator network coupling spatial and spectral regularization for geometric consistency. \\
As our model uses a vectorized representation of the target motion, and for fair comparison, we train the other models to reconstruct the whole sequence. ScanMove outperforms the aforementioned state-of-the-art models for the MSE metric and comes close second for the Cosim metric. To further support these results, we show a qualitative comparison in~\cref{fig:qualitative_comparisons_DFAUST}.

\begin{table}[ht!]
\centering
\begin{tabular}{lcc}
\toprule
         & MSE $\downarrow$ & Cosim $(\times 10^{-1}) \downarrow$ \\ \midrule

NJF \cite{NJF_2022}     & 9.50  &   1.30   \\
Meshconv \cite{Meshconv_2020} & 6.62  &   1.42  \\
ARAPReg \cite{ARAPReg}  & 5.49   &  1.56  \\
SMS \cite{SMS_2024}    & 4.42  &   \textbf{1.13}  \\ \midrule
Ours    &  \textbf{ 4.23}  &  1.20  \\
\bottomrule
\end{tabular}
\caption{\textbf{Metrics for shape interpolation on DFAUST.} We report the mean squared error between predicted vertex positions and ground truth positions along with mean cosine similarity between predicted normals and ground truth normals.}
\end{table}

Graph convolutional models (ARAPReg and Meshconv) struggle to reconstruct regions with small triangles such as hands, even with ARAP regularization: we believe this is due to the low amount of training samples in our training dataset compared to the one in the original papers. Similarly, NJF learns deformations from a PoinNet global encoder that shows poor efficiency when training data is scarce. Finally, SMS shows satisfying performance but was observed to struggle with the original DFAUST meshing which is non-isotropic making spectral regularization harder.

\subsection{Walking and running motion generation}
\begin{figure}[ht!]
    \centering
    \includegraphics[width=0.8\linewidth]{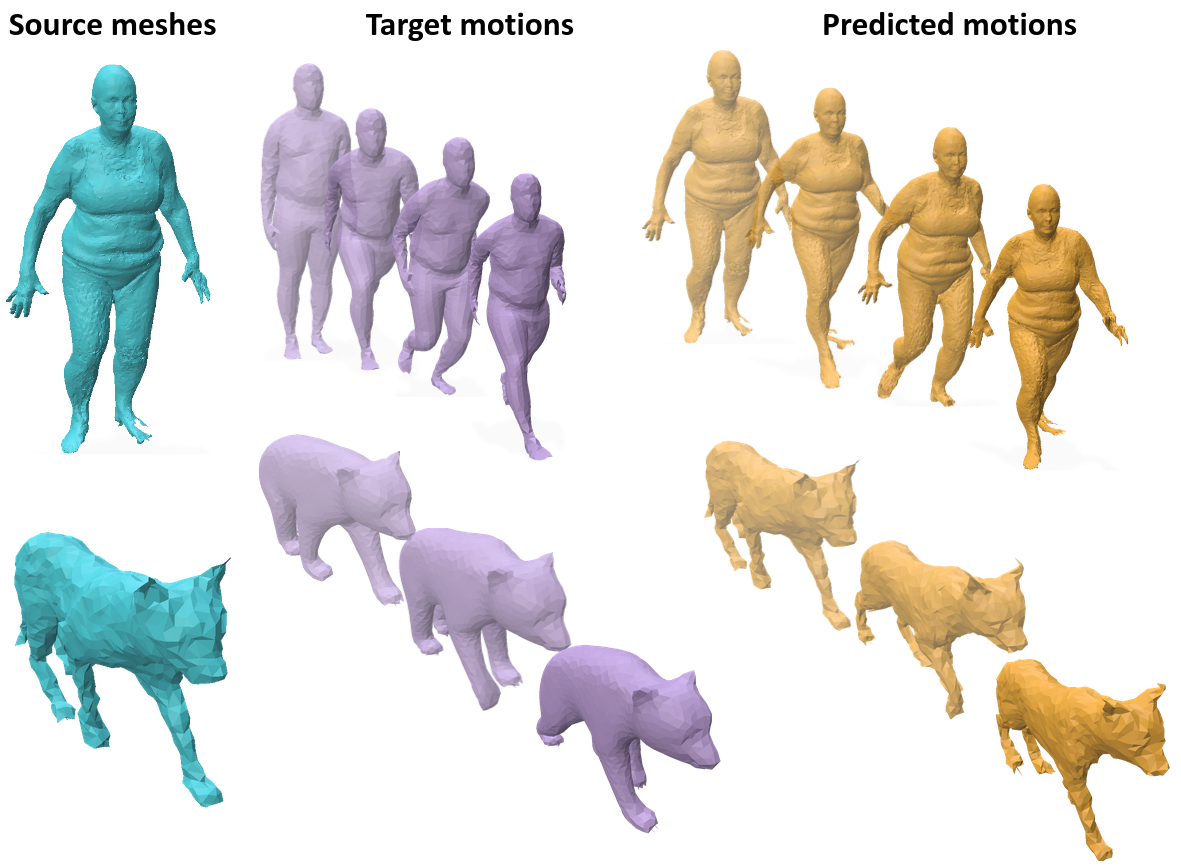}
    \caption{\textbf{Examples of motion transfer on different source meshes.} We transfer the encoded motions (\textcolor{violet}{in purple}) to the unregistered source meshes (\textcolor{cyan}{in blue)} and generate the deformations (\textcolor{orange}{in orange}). On the top row is an example with a scan from FAUST and on the bottom row is an example with the animal category of DeformingThings4D.}
    \label{fig:motion_transfer_ex}
\end{figure}

\begin{figure*}[ht!]
    \centering
    \includegraphics[width=1.0\linewidth]{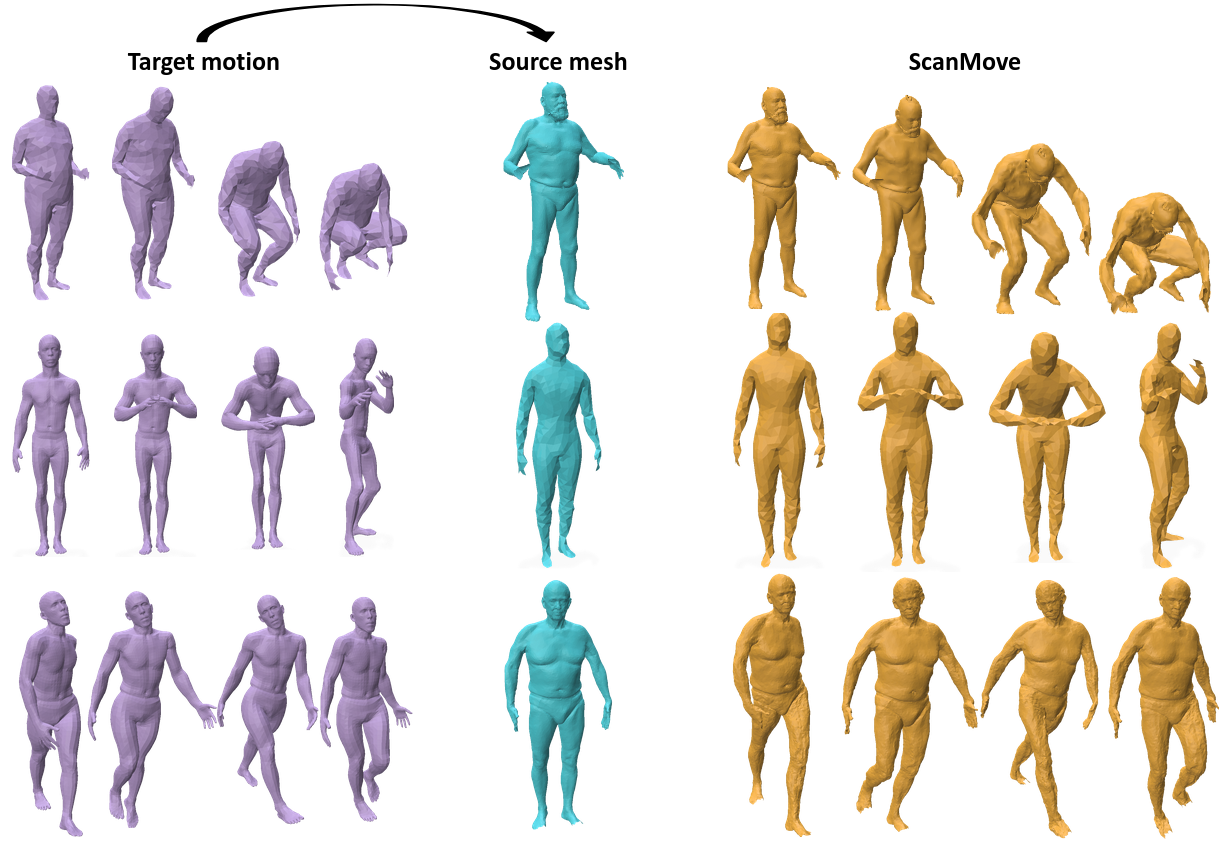}
    \caption{\textbf{Examples of motion transfer on different source meshes.} We transfer the encoded motions (\textcolor{violet}{in purple}) to the unregistered source meshes (\textcolor{cyan}{in blue)} and generate the deformations (\textcolor{orange}{in orange}).}
    \label{fig:motion_transfer_more}
\end{figure*}

Next, we evaluate our method for "long-range" motions such as walking. The test set comprises two identities walking in three different directions. Few comparable rig-free and correspondence-free methods are available: we modified the existing NJF \cite{NJF_2022} and 3DCODED \cite{groueix2018_3DCODED} models to predict successive deformations of the source mesh to predict a whole deformation sequence. 

\begin{table}[ht!]
\centering
\begin{tabular}{lcc}
\toprule
& MSE $\downarrow$ & Cosim $(\times 10^{-2})\downarrow$ \\ \midrule
3DCODED \cite{groueix2018_3DCODED} &  12.45   &   9.03   \\
NJF \cite{NJF_2022} &   6.26  &     \textbf{4.20}    \\ \midrule
Ours  &  \textbf{4.24}   &   5.10  \\
\bottomrule
\end{tabular}
\caption{\textbf{Metrics for motion generation on WALK.} We report the mean squared error between predicted vertex positions and ground truth positions along with mean cosine similarity between predicted normals and ground truth normals.}
\end{table}

Once again, ScanMove showed better performance with respect to the MSE metric. The Cosim metric is better for NJF as it is a first-order model in the sense that it has an inherent smoothing coming from solving the Poisson equation which smooth out the deformation and so does the normal map of the deformed mesh. We also show qualitative comparison in~\cref{fig:qualitative_comparisons_DFAUST}.
Overall, ScanMove performs better with motion sequences thanks to its "time-awareness" arising from the motion embedding and its recursive design acting on learned spatiotemporal features.

\subsection{Applications}
\label{subsec:applications}
\begin{figure}[ht!]
    \centering
    \includegraphics[trim=130 0 170 0,clip,width=1.0\linewidth]{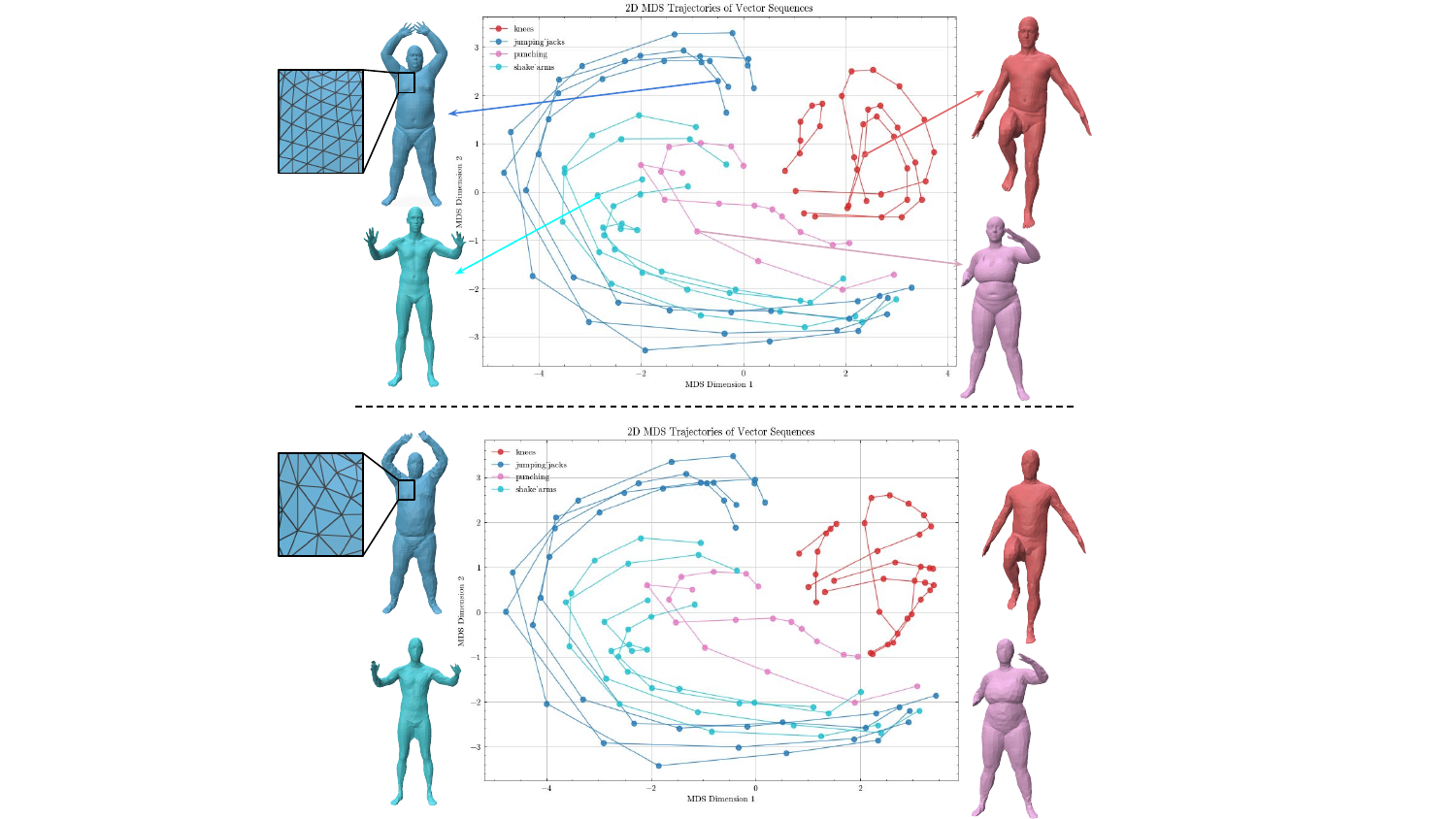}
    \caption{\textbf{Multi-dimensional scaling (MDS) of encoded motions from DFAUST.} We compute the MDS of four motions from DFAUST obtained with the original face connectivity (top). We compare with the same MDS on a low-resolution connectivity (bottom).}
    \label{fig:mds_motion_embedding}
\end{figure}

\noindent \textbf{Rig-free motion transfer.} In a similar way to \cite{TRJ}, the proposed framework enables rig-free motion transfer. Some examples are showcased in the supplementary material videos, and we display a series of examples in~\cref{fig:motion_transfer_ex} and~\cref{fig:motion_transfer_more}. Note that, as shown in the figure, ScanMove is versatile enough to be adapted to different challenging shape categories such as scans from the FAUST dataset \cite{FAUST_2014} and animals from the DeformingThings4D dataset \cite{DeformingThings4D}.

\noindent \textbf{Motion embedding.} In addition to the prediction of time-dependent deformations, the proposed motion embedding network can be used once trained for other tasks. By design, the predicted vector representation of motions is robust against remeshing. To support this, in~\cref{fig:mds_motion_embedding}, we compute a multi-dimensional scaling (MDS) of the embedding vectors onto a 2D space and we do this for the original DFAUST parameterization (top row) and for a downsampled version (bottom row). Each point trajectory in the graphs corresponds to a motion. We highlight that the two graphs are almost identical and that similar movements share similar patterns in the embedding space: see for example the blue curves compared to the cyan curves which respectively correspond to raising arms above the head and raising arms in front of the body. On the other hand, raising a knee (in red) is clearly separated from the other two motions mentioned.

\subsection{Ablation}
To support our design choice, in~\cref{tab:ablation_studies}, we report the performance metrics on the WALK dataset with different design variations.

\begin{table}[ht!]
\centering
\begin{tabular}{llcc}
\toprule 
 &     & \textbf{MSE} $\downarrow$ & \textbf{Cosim} $\downarrow$ \\
\midrule
\multirow{1}{*}{\textit{Motion embed.}} & /wo GRU  &  6.35  & 7.12  \\
\midrule
\multirow{3}{*}{\textit{Loss}}  & $\mathcal{L}^{MSE}$ & 4.89   &  5.98   \\
& $\mathcal{L}^{MSE}$ + $\mathcal{L}^N$ &  4.37  &  5.14   \\
                         &  $\mathcal{L}^{CH}$ & 8.35 &  8.14   \\
\midrule
\textbf{Ours} &  (GRU + $\mathcal{L}$)   &  \textbf{4.24}  &  \textbf{5.10}  \\
\bottomrule
\end{tabular}
\caption{\textbf{Ablation studies evaluated on the WALK dataset.} We report MSE and cosine similarity when replacing the model of the modules (1st column) with other candidates (2nd column).}
\label{tab:ablation_studies}
\end{table}

In a complete unsupervised setting (using $\mathcal{L}^{CH}$), performance are not as good as in the registered setting (see \cref{tab:ablation_studies}) but to our knowledge, this is the first method to address complete unsupervised learning of complex motion prediction.

We also tested multiple candidates for the different component of the framework, namely the feature extractor, the motion frame encoder, the time consistency module and the deformation generator. While one could expect more advanced models such as Prism \cite{attaiki2023snk}, transformers \cite{transformers_2017} or NJF \cite{NJF_2022} to show better performance, we found these model tricky to optimize correctly or prohibitively expensive for long range motions. 

\subsection{Computation time}
Once trained, a major advantage for ScanMove compared to optimization methods is fast inference and a linear time complexity with respect to the resolution of the input mesh, defined as the number of vertices of the source mesh $\# V_{source}$. We report comparisons with an optimization based method \cite{deformation_transfer_2004} and a state-of-the-art robust deep deformation model \cite{SMS_2024} in~\cref{tab:inference_time}.

\begin{table}[ht!]
\small
\centering
\begin{tabular}{lccc}
\toprule
\# $V_{source}$ &  low (2000) &  mid (15000) & high (50000)\\ \midrule
\cite{deformation_transfer_2004}  &  32  &  243  &  768  \\
SMS \cite{SMS_2024}  &  16  &  28  &  99  \\ \midrule
Ours  &  \textbf{7}  &  \textbf{18} & \textbf{64} \\
\bottomrule
\end{tabular}
\caption{\textbf{Inference time (in seconds) to generate a whole sequence of 200 frames.} We report the mean over 3 inferences. For \cite{deformation_transfer_2004}, we only reported the time for deformation transfer but the method also require a dense correspondence estimation step.}
\label{tab:inference_time}
\end{table}

\subsection{Remeshing robustness}
Robustness to a change of discretization is offered thanks to the robustness of each component of the framework. We evaluate said robustness on different remeshing of the DFAUST test set. In particular, we report the performance drop when evaluating our model on the original mesh parameterization (Original), a factor 2 downsampling (DS(2)), a factor 2 upsampling (US(2)) and variable density (VD) for which one half of the mesh is upsampled and the other half is downsampled. We define the relative difference $\Delta_{rel}$ of the metric $Met$ between the original parameterization $Orig$ and the corresponding remeshing $Remesh$.

\begin{equation}
    \Delta_{rel} = \frac{|Met(Remesh) - Met(Orig)|}{Met(Orig)}
\end{equation}

\begin{figure}[ht!]
    \centering
    \includegraphics[width=1.0\linewidth]{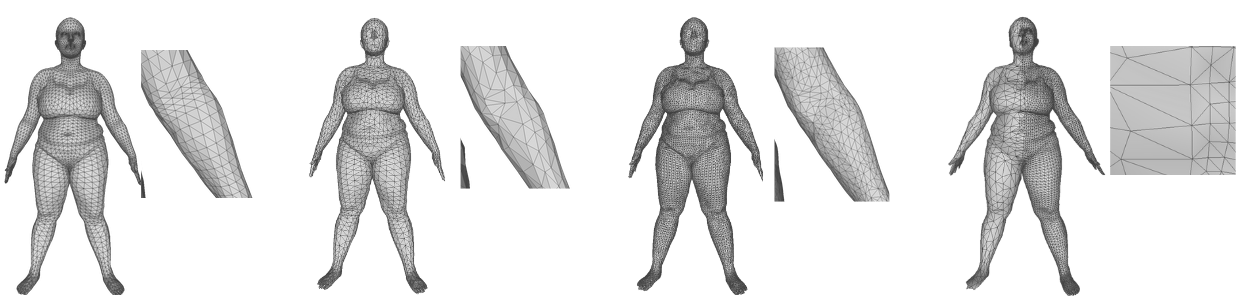}
    \caption{\textbf{Different discretization used for the robustness experiment. From left to right: original topology, DS(2), US(2) and variable density (VD).}}
    \label{fig:robustness_exp_setting}
\end{figure}

\begin{table}[ht!]
\centering
\begin{tabular}{lcc}
\toprule
 & \textbf{MSE deviation} & \textbf{Cosim deviation} \\
\midrule
Original  &  0\% & 0\%  \\
DS(2)  &  3.9\%  & 1.2\%  \\
US(2)  &  4.1\%  & 3.1\%  \\
VD     &  12.1\% & 8.3\% \\
\bottomrule
\end{tabular}
\caption{\textbf{Metrics deviation from the training mesh topology for different types of remeshing of the test set.} We report the relative difference (in \%)}
\label{tab:robustness_results}
\end{table}

\noindent Results are reported in \ref{tab:robustness_results}.With less than 5\% in performance drop from the original training discretization, we validate the robustness of the proposed framework against changes in the mesh topology.

\subsection{Tests with other models}
Here, we report our test results using other candidates for the different parts of our framework. The numbers are obtained by training and testing the framework after replacing the designated component (first column) with the proposed model (second column) on the WALK dataset.

\begin{table}[ht!]
\small
\centering
\begin{tabular}{llcc}
\toprule 
 &     & \textbf{MSE} $\downarrow$ & \textbf{Cosim} $(\times 10^{-2})\downarrow$ \\
\midrule
\multirow{1}{*}{\textit{Motion emb.}} & SA \cite{transformers_2017} &  9.59  &  10.35  \\
\midrule
\multirow{1}{*}{\textit{Feat. ext.}} & MLP  & 16.36  & 14.9  \\
\midrule
\multirow{2}{*}{\textit{Decoder}}  &  Prism \cite{attaiki2023snk}  &  7.21  &  8.54    \\
     & NJF \cite{NJF_2022}     &  5.07  & 6.06  \\
\midrule
\textbf{Ours} &     &  \textbf{4.24}  &  5.10  \\
\bottomrule
\end{tabular}
\caption{\textbf{Tests reports on the WALK dataset.} We report MSE and cosine similarity when replacing the model of the modules (1st column) with other candidates (2nd column).}
\label{tab:tests_other_candidates}
\end{table}

\noindent We observed worse results with more advanced models. For the self-attention (SA), this could be due to the recursive aspect in which we do not require long range dependencies. For the decoder, the tested models \cite{NJF_2022, attaiki2023snk} have not been used in recursive frameworks before and they might require more stability to train properly.

\section{Limitations and future work}
\label{sec:limitations}
This work opens up interesting avenues for future research. For example, while our motion embedding network based on PointNet effectively captures extrinsic changes like pose variations addressed in this paper, integrating it with a complementary fine-grained encoder could strengthen its ability to model non-isometric, intrinsic deformations. Additionally, although the recursive deformation generator delivers coherent sequences, introducing corrective feedback mechanisms would make it even more robust for very long motions. In this work, we relied on the ability of the network to self correct during the training process but developing new approaches for periodic resetting of the deformation could be an interesting idea to mitigate error accumulation over time.

\section{Conclusion}
\label{sec:conclusion}

In this paper, we introduced ScanMove, a novel framework for predicting complex, time-dependent deformations of unregistered body meshes without reliance on skeletal rigging or explicit alignment in training data. Our method successfully couples a robust mesh feature extractor with a motion embedding network and a recursive deformation generator. Extensive experiments demonstrate that ScanMove predicts realistic deformations in various challenging scenarios. The ability of the method to capture both short- and long-range deformations, along with its versatility in applications like motion transfer and unsupervised learning, underlines its potential for advancing motion synthesis in fields like animation, environment modeling, and robotics. Overall, ScanMove offers strong foundations for future work in complex unsupervised settings.


\section*{Acknowledgments}

This work benefited from the support of the project project 4DSHAPE ANR-24-CE23-5907 of the French National Research Agency (ANR).







\end{document}